\documentclass{siamltex}

\usepackage{epsfig}

\begin{document}

\pagestyle{myheadings} \markboth{\sc he, kao, and lu}{\sc linear-time
succinct encodings of planar graphs}

\title{Linear-Time Succinct Encodings of Planar Graphs \\ via Canonical
Orderings}

\author{ 
Xin He\thanks{Department of Computer Science and Engineering, State
University of New York
at Buffalo, Buffalo, NY 14260.  Email: xinhe@cse.buffalo.edu.  Research
supported in part by NSF Grant CCR-9205982.}
\and 
Ming-Yang Kao\thanks{
Department of Computer Science, Yale University,
New Haven, CT 06250, USA.  Email: kao-ming-yang@cs.yale.edu.  Research
supported in part by NSF Grant CCR-9531028.}
\and
Hsueh-I Lu\thanks{Department of Computer Science and Information
Engineering, National Chung-Cheng University, Chia-Yi 621,
Taiwan, ROC. Email: hil@cs.ccu.edu.tw.}}

\maketitle

\begin{abstract} 
Let $G$ be an embedded planar undirected graph that has $n$ vertices,
$m$ edges, and $f$ faces but has no self-loop or multiple edge. If $G$
is triangulated, we can encode it using $\frac{4}{3}m-1$ bits,
improving on the best previous bound of about $1.53m$ bits.  In case
exponential time is acceptable, roughly $1.08m$ bits have been known
to suffice. If $G$ is triconnected, we use at most
$(2.5+2\log{3})\min\{n,f\}-7$ bits, which is at most $2.835m$ bits and
smaller than the best previous bound of $3m$ bits.  Both of our
schemes take $O(n)$ time for encoding and decoding.
\end{abstract} 



\section{Introduction} This paper investigates the problem of {\it
encoding} a given graph $G$ into a binary string $S$ with the requirement
that $S$ can be {\it decoded} to reconstruct $G$.  The problem has been
studied generally with two primary objectives. One is to minimize the
length of $S$, while the other is to minimize the time needed to compute
and decode $S$. In light of these goals, a coding scheme is {\em
efficient} if its encoding and decoding procedures both take polynomial
time.  A coding scheme is {\em succinct} if the length of $S$ is not much
larger than its {\em information-theoretic tight bound}, i.e., the shortest
length over all possible coding schemes.

As the two primary objectives are often in conflict, a number of
coding schemes with different trade-offs have been proposed from
practical and theoretical perspectives.  The most well-known efficient
succinct scheme is the folklore scheme of encoding a rooted ordered
$n$-vertex tree into a string of balanced $n-1$ pairs of left and
right parentheses, which uses $2(n-1)$ bits. Since the total number of
such trees is at least
$\frac{1}{2(n-1)}{\cdot}\frac{(2n-2)!}{(n-1)!(n-1)!}$, the minimum
number of bits needed to differentiate these trees is the
logarithm\footnote{All logarithms are of base 2.} of this quantity,
which is $2n-o(n)$ by Stirling's approximation.  Thus, 2 bits per edge
is an information-theoretic tight bound for encoding rooted ordered
trees.  The standard adjacency-list encoding of a graph is widely
useful but requires $\Theta(m{\log} n)$ bits where $m$ and $n$ are the
numbers of edges and vertices, respectively~\cite{clr}.  For certain
graph families, Kannan, Naor and Rudich~\cite{KNR92} gave schemes that
encode each vertex with $O(\log n)$ bits and support $O(\log n)$-time
testing of adjacency between two vertices.  For connected planar
graphs, Jacobson \cite{Jacobson89} gave an $\Theta(n)$-bit encoding
which supports traversal in $\Theta(\log{n})$ time per vertex visited.
This result was recently improved by Munro and Raman \cite{MR97};
their schemes encode binary trees, rooted ordered trees and planar
graphs succinctly and support several graph operations in constant
time.  For dense graphs and complement graphs, Kao, Occhiogrosso, and
Teng~\cite{kaot93.joa} devised two compressed representations from
adjacency lists to speed up basic graph techniques such as
breadth-first search and depth-first search.  Galperin and Wigderson
\cite{GW83} and Papadimitriou and Yannakakis~\cite{PH86.encode}
investigated complexity issues arising from encoding a graph by a
small circuit that computes its adjacency matrix.  For labeled planar
graphs, Itai and Rodeh \cite{IR82} gave an encoding procedure that
requires $\frac{3}{2} n \log n + O(n)$ bits.  For unlabeled general
graphs, Naor \cite{naor90} gave an encoding of
$\frac{n^2}{2}-n\log{n}+ O(n)$ bits, which is optimal to the second
order.

Our work aims to minimize the number of bits needed to encode an embedded
planar graph $G$ which is unlabeled and undirected.  We assume that $G$ has
$n$ vertices, $m$ edges, and $f$ faces but has no self-loop or multiple
edge. (See \cite{Ber85, Har72, Lovasz_book_86} for the graph-theoretic
terminology used in this paper.)  Note that if polynomial time for encoding
and decoding is not required, then any given graph in a large family can be
encoded with the information-theoretic minimum number of bits by
brute-force enumeration.  This paper focuses on schemes that use only
$O(n)$ time for both encoding and decoding.

For a general planar graph $G$, Tur\'{a}n \cite{turan84} gave an encoding
using $4m$ bits asymptotically. This space complexity was improved by
Keeler and Westbrook \cite{KW:encodings} to about $3.58m$ bits.  They also
gave encoding algorithms for several important classes of planar graphs. In
particular, they showed that if $G$ is triangulated, it can be encoded in
about $1.53m$ bits. If $G$ is triconnected, it can be encoded using $3m$
bits. In this paper, these latter two results are improved as follows. If
$G$ is triangulated, it can be encoded using $\frac{4}{3}m-1$ bits. It is
interesting that rooted ordered trees require 2 bits per edge, while the
seemingly more complex plane triangulations need fewer bits.  Note that
Tutte~\cite{Tutte62} gave an enumeration theorem that yields an
information-theoretic tight bound of roughly $1.08m$ bits for plane
triangulations that may contain multiple edges.  If $G$ is triconnected, we
can encode it using at most $(2.5+2\log{3})\min\{n,f\}-7$ bits, which is at
most $2.835m$ bits.  Both of our coding schemes are intuitive and
simple. They require only $O(n)$ time for encoding as well as decoding.
The schemes make new uses of the canonical orderings of planar
graphs, which were originally introduced by de Fraysseix, Pach and
Pollack \cite{DeFPP90} and extended by Kant \cite{Kant92}. These structures
and closely related ones have proven useful also for drawing planar graphs
in organized and compact manners \cite{KH97,kaofhr94,Schnyder90}.

This paper is organized as follows. In \S\ref{sec_triangle}, we present our
coding scheme for plane triangulations. In \S\ref{sec_triconnect}, we
generalize the scheme to encode triconnected plane graphs. We conclude the
paper with some open problems in \S\ref{sec_open}.

\section{A Coding Scheme for Plane Triangulations}\label{sec_triangle} 

This section assumes that $G$ is a plane triangulation.  Thus, $n \geq 3$
and $G$ has $m=3n-6$ edges.

Let $v_1,\ldots,v_n$ be an ordering of the vertices of $G$, where
$v_1,v_2,v_n$ are the three exterior vertices of $G$ in the
counterclockwise order.  After fixing such an ordering, let $G_k$ be
the subgraph of $G$ induced by $v_1,\ldots,v_k$. Let $H_k$ be the
exterior face of $G_k$. Let $G-G_k$ be the subgraph of $G$ obtained by
removing $v_1,\ldots,v_k$.
Our coding scheme uses a special kind of ordering defined as follows.
\begin{definition}[see \cite{DeFPP90}]\rm\label{canonical}
An ordering $v_1,\ldots,v_n$ of $G$ is {\em canonical} if the
following statements hold for every $k=3,\ldots,n$:

\begin{enumerate}
\item  $G_k$ is biconnected,  and its exterior face $H_k$ is a cycle
containing the edge $(v_1,v_2)$.

\item The vertex $v_k$ is on the exterior face of $G_k$, and  the set of its
neighbors in $G_{k-1}$ forms a subinterval of the path
$H_{k-1}-\{(v_1,v_2)\}$ and consists of at least two vertices. Furthermore,
if $k < n$, $v_k$ has at least one neighbor in $G-G_k$.  Note that the case
$k=3$ is somewhat ambiguous due to degeneracy, and $H_2-\{(v_1,v_2)\}$ is
regarded as the edge $(v_1,v_2)$ itself.
\end{enumerate}
\end{definition}

\begin{figure}[t]
\centerline{\psfig{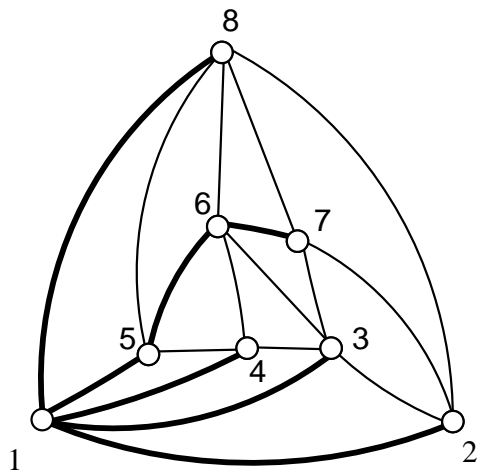}}
\caption{A plane triangulation and a canonical ordering.}
\label{canonical-triangulate}
\end{figure}

Figure~\ref{canonical-triangulate} illustrates a canonical ordering of
a plane triangulation.  Note that every plane triangulation has a
canonical ordering which can be computed in $O(n)$ time
\cite{DeFPP90}.  A canonical ordering of $G$ can be viewed as an order
in which $G$ is reconstructed from a single edge $(v_1,v_2)$ step by
step. At step $k$ with $3 \leq k \leq n$, the vertex $v_k$ and the
edges between $v_k$ and its lower ordered neighbors are added into the
graph.  For the sake of enhancing intuitions, we call $H_{k-1}$ the
{\em contour} of $G_{k-1}$; denote its vertices by
$c_1(=v_1),c_2,\ldots,c_{t-1},c_t(=v_2)$ in the consecutive order
along the cycle $H_{k-1}$; and visualize them as arranged from left to
right above the edge $(v_1,v_2)$ in the plane.  When the vertex $v_k$
is added to $G_{k-1}$ to construct $G_k$, let
$c_\ell,c_{\ell+1},\ldots,c_r$ be the neighbors of $v_k$ on the
contour $H_{k-1}$. After $v_k$ is added, the vertices
$c_{\ell+1},\ldots,c_{r-1}$ are no longer contour vertices. Thus, we
say that these vertices are {\em covered} by $v_k$. The edge
$(v_k,c_\ell)$ is the {\em left} edge of $v_k$; the edge $(v_k,c_r)$
is the {\em right} edge of $v_k$; the edges $(c_p,v_k)$ with $\ell < p
< r$ are the {\em internal} edges of $v_k$.

There is no published reference for the following folklore lemma; for
the sake of completeness, we include its proof here.

\begin{lemma}\label{lemma:3-tree}
Let $v_1,\ldots,v_n$ be a canonical ordering of $G$.  Let $T_1$
$($respectively, $T_2$$)$ be the collection of the left
$($respectively, right$)$ edges of $v_j$ for $3 \leq j \leq n-1$;  
similarly, let $T_n$ be that of the internal edges of $v_j$ for $3
\leq j \leq n$.
\begin{enumerate}
\item\label{lemma:3-tree_1}
$T_1$ is a tree spanning over $G-\{ v_2,v_n\}$.
\item\label{lemma:3-tree_2}
$T_2$ is a tree spanning over $G-\{ v_1,v_n\}$.
\item\label{lemma:3-tree_3}
 $T_n$ is a tree spanning over $G-\{ v_1,v_2\}$.
\end{enumerate}
\end{lemma}

\begin{proof} The statements are proved separately as follows.

Statement \ref{lemma:3-tree_1}. For $i=3,\ldots,n-1$, let $D_i$ be
the collection of the left edges of $v_j$ for $3 \leq j \leq i$.  We
prove by induction on $i$ the claim that $D_i$ is a tree spanning over
$v_1,v_3,\ldots,v_i$. Then, since $T_1=D_{n-1}$, the claim implies the
statement. For the base case $i=3$, the claim trivially holds. The
induction hypothesis is that the claim holds for $i=k-1<n-1$.  The
induction step is to prove the claim for $i=k \leq n-1$. $D_k$ is
obtained from $D_{k-1}$ by adding the left edge $(v_k,c_\ell)$ of
$v_k$.  By the induction hypothesis, $D_{k-1}$ is a tree spanning over
$v_1,v_3,\ldots,v_{k-1}$.  Since $c_\ell$ is the leftmost neighbor of
$v_k$ on $H_{k-1}$, $c_\ell$ is some $v_j$ with $1 \leq j \leq k-1$
and $j \not= 2$.  Thus, $D_{k-1}$ contains $c_\ell$, and $D_k$ is a
tree spanning over $v_1,v_3,\ldots,v_{k-1},v_k$.

Statement \ref{lemma:3-tree_2}. The proof is symmetric to that of
Statement \ref{lemma:3-tree_1}.

Statement \ref{lemma:3-tree_3}. $G$ has $n$ vertices and $3n-6$ edges.
The edges $(v_1,v_2),(v_2,v_n),(v_1,v_n)$ are not in $T_1 \cup T_2
\cup T_n$.  Thus, since $T_1$ and $T_2$  have $n-3$ edges each, $T_n$
has $n-3$ edges. Then, since $T_n$ is acyclic and does not contain
$v_1$ and $v_2$, $T_n$ is a spanning tree of $G-\{v_1,v_2\}$.
\end{proof}

A canonical ordering $v_1,\ldots,v_n$ is {\em rightmost} if for all
$v_k$ and $v_{k'}$ with $k' > k$ such that the neighbors of $v_{k'}$
on $H_{k'-1}$ are all in $H_{k-1}$, the leftmost neighbor of $v_{k'}$
appears before that of $v_k$ when traversing $H_{k-1}$ from $v_1$ to
$v_2$ in the clockwise direction.  Intuitively speaking, if there are
more than one vertex that can be added to $G_{k-1}$, we always add the
rightmost one.  The ordering in Figure~\ref{canonical-triangulate} is
rightmost.  A rightmost canonical ordering is symmetric to a {\em
leftmost} one in \cite{Kant92} and can be computed from $G$ in linear
time similarly.

Let $v_1,\ldots,v_n$ be a rightmost canonical ordering of $G$.  Let
$T_1$ be as in Lemma \ref{lemma:3-tree} for this ordering. Let $T$ be
the tree $T_1 \cup \{(v_1,v_n), (v_1,v_2)\}$.  In
Figure~\ref{canonical-triangulate}, $T$ is indicated by the thick
lines.  Our coding scheme uses $T$ extensively.  The {\em rightmost}
depth-first search of $T$ proceeds as follows. We start at $v_1$ and
traverse the edge $(v_1,v_2)$ first. Afterwards, if two or more
vertices can be visited from $v_k$, we choose the rightmost one. More
precisely, let $P$ be the path in $T$ from $v_k$ to $v_1$ and then to
$v_2$.  Let $D$ be the set of edges between $v_k$ and the available
vertices. We visit a new vertex through the edge in $D$ that is next
to $P$ in the counterclockwise cyclic order around $v_k$ formed by $P$
and the edges in $D$.  Note that the order in which the vertices are
visited by the rightmost depth-first search is the rightmost canonical
ordering $v_1,\ldots,v_n$ that defines $T$.

We are now ready to describe the encoding $S$ of $G$ as the concatenation
of two binary strings $S_1$ and $S_2$ as follows.

$S_1$ is the binary string that encodes $T$ using the folklore parenthesis
coding scheme where $0$ and $1$ correspond to $''(''$ and $'')''$,
respectively.  In this encoding, $T$ is rooted at $v_1$, and the branches
are ordered the same as their enpoints are in the rightmost canonical
ordering. Since $T$ contains $n$ vertices, $S_1$ has $2(n-1)$ bits.

$S_2$ encodes the number of contour vertices covered by each $v_k$
with $3\leq k\leq n$. First, we create a string of $n-2$ copies of
$0$. The $(k-2)$-th $0$ corresponds to $v_k$.  If $v_k$ covers $d$
vertices, we insert $d$ copies of $1$ before the corresponding $0$.
For example, the string $S_2$ for Figure \ref{canonical-triangulate}
is:
\[00010101110\]
Since each vertex $v_k$ with $3\leq k\leq n-1$ is covered exactly once, $S_2$
has $n-3$ copies of 1. So $|S_2|=(n-2)+(n-3)=2n-5$ bits.  Hence,
$|S|=|S_1|+|S_2|=4n-7$ bits.

We next describe how to decode $S$ to reconstruct $G$.  Given $S$, we can
uniquely determine $n$ from the length of $S$.  Subsequently, we can
uniquely determine $S_1$ and $S_2$.  {From} $S_1$, we can reconstruct $T$.
{From} $T$, we can recover the ordering $v_1,\ldots,v_n$.  Then, we draw
the edge $(v_1,v_2)$ and perform a loop of $n-2$ steps indexed by $k$ with $3
\leq k \leq n$ where step $k$ processes $v_k$. Before $v_k$ is processed,
$G_{k-1}$ and its contour $H_{k-1}$ have been constructed.  At step $k$, we
add $v_k$ and the edges between $v_k$ and its lower ordered neighbors into
$G_{k-1}$ to construct $G_k$ as follows.  {From} $T$, we can identify the
leftmost neighbor $c_\ell$ of $v_k$ on the contour $H_{k-1}$, because $c_\ell$ is
simply the parent of $v_k$ in $T$.  {From} $S_2$, we can determine the
number $d$ of vertices covered by $v_k$. Thus, we add the edges
$(c_\ell,v_k),(c_{\ell+1},v_k),\ldots,(c_{\ell+d+1},v_k)$ into $G_{k-1}$; note that
$r = \ell+d+1$.  This gives us the subgraph $G_k$ and completes step $k$.

It is straightforward to carry out these encoding and decoding procedures
in linear time. Also, we can save 1 bit by deleting the last $0$ in $S_2$.
Since $v_3$ covers no vertex, for $n \geq 4$, we can save another bit by
deleting the first $0$ in $S_2$.  Note that for $n=3$, the last $0$ in
$S_2$ is also the first $0$ and cannot be deleted twice, but we can simply
encode the 3-vertex plane triangulation with zero bit without ambiguity.
Thus, we have the following theorem.

\begin{theorem}\label{encode-triangulation}
A plane triangulation of $m$ edges and $n$ vertices with $n \geq 4$
can be encoded using $4n-9=\frac{4}{3}m-1$ bits. Both encoding and
decoding take $O(n)$ time.
\end{theorem}

\section{A Coding Scheme for Triconnected Plane Graphs} 
\label{sec_triconnect}
This section assumes that $G$ is triconnected. To avoid triviality,
let $n \geq 3$.

Let $v_1,\ldots,v_n$ be an ordering of the vertices of $G$ where
$v_1,v_2,v_n$ are on the exterior face of $G$, and $v_2$ and $v_n$ are
neighbors of $v_1$.  Let $G_k$ be the subgraph of $G$ induced by
$v_1,\ldots,v_k$. Let $H_k$ be the exterior face of $G_k$.  Let
$G-G_k$ be the subgraph of $G$ obtained by removing $v_1,\ldots,v_k$.
Our coding scheme for triconnected plane graphs uses an ordering
defined as follows.

\begin{definition}[see \cite{Kant92}]\label{def_kant}\rm
An ordering $v_1,\ldots,v_n$ of a triconnected plane graph $G$ is {\em
canonical} if the integer interval $[3,n]$ can be partitioned into
subintervals $[k,k+q]$ each satisfying either set of properties below:
\begin{enumerate}
\item The integer $q$ is $0$.  The vertex $v_k$ is on the exterior face of $G_k$
and has at least two neighbors in $G_{k-1}$. $G_k$ is biconnected and its
exterior face contains the edge $(v_1,v_2)$.  If $k < n$, $v_k$ has at least
one neighbor in $G-G_k$.

\item The integer $q$ is at least $1$. The sequence
$v_k,v_{k+1},\ldots,v_{k+q}$ is a chain on the exterior face of $G_{k+q}$ and
has exactly two neighbors in $G_{k-1}$, one for $v_k$ and the other for
$v_{k+q}$, which are on the exterior face of $G_{k-1}$. $G_{k+q}$ is
biconnected and its exterior face contains the edge $(v_1,v_2)$. Every vertex
among $v_k,\ldots,v_{k+q}$ has at least one neighbor in $G-G_{k+q}$.
\end{enumerate}
\end{definition}

\begin{figure}[t]
\centerline{\psfig{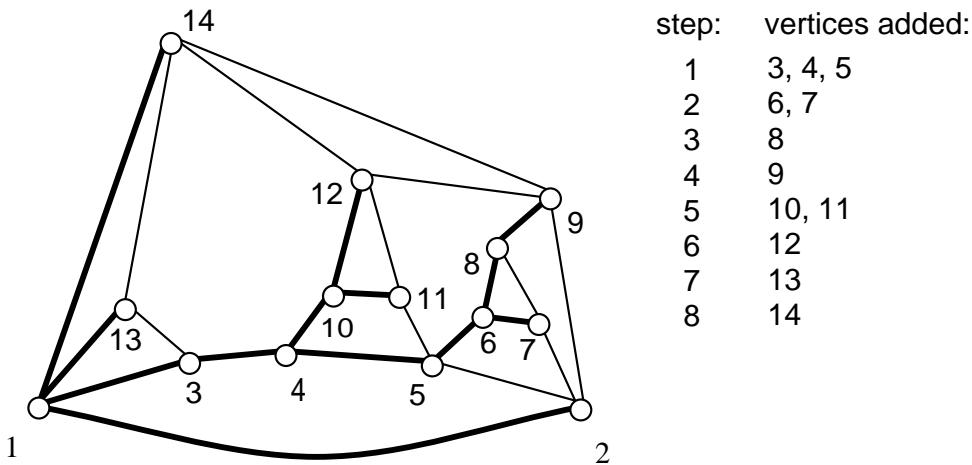}}
\caption{A triconnected plane graph and a canonical ordering.}
\label{canonical-triconnected}
\end{figure}

As in \S\ref{sec_triangle}, we similarly define a {\em rightmost}
canonical ordering $v_1,\ldots,v_n$ of
$G$. Figure~\ref{canonical-triconnected} shows a rightmost canonical
ordering of a triconnected plane graph. Given a triconnected plane
graph, we can find a rightmost canonical ordering in linear
time~\cite{Kant92}. With a rightmost canonical ordering, $G$ can be
reconstructed from a single edge $(v_1,v_2)$ through a sequence of
steps indexed by $k'$. There are two possible cases at step $k'$,
which correspond to the two sets of properties in
Definition~\ref{def_kant} and are used throughout this section.

{\em Case 1}: A single vertex $v_k$ is added.  

{\em Case 2}: A chain of $q+1$ vertices $v_k,\ldots,v_{k+q}$ is added.

While reconstructing $G$, we collect a set $T$ of edges as
follows. Initially, $T$ consists of the edge $(v_1,v_2)$. Let
$c_1(=v_1),c_2,\ldots,c_{t-1},c_t(=v_2)$ be the vertices of $H_{k-1}$,
which are ordered consecutively along the boundary cycle of $H_{k-1}$
and are arranged from left to right above the edge $(v_1,v_2)$ in the
plane.

Case 1. Let $c_\ell$ and $c_r$ with $1\leq \ell < r \leq t$ be the
leftmost and rightmost neighbors of $v_k$ in $H_{k-1}$,
respectively. After $v_k$ is added, $c_{\ell+1},\ldots,c_{r-1}$ are no
longer contour vertices; these vertices are {\em covered} at step
$k'$. The edge $(c_\ell,v_k)$ is included in $T$.

Case 2. Let $c_\ell$ and $c_r$ with $1\leq \ell < r \leq t$ be the
neighbors of $v_k$ and $v_{k+q}$ in $H_{k-1}$, respectively. After
$v_k,\ldots,v_{k+q}$ are added, $c_{\ell+1},\ldots,c_{r-1}$ are no
longer contour vertices; these vertices are {\em covered} at step
$k'$. The edges
$(c_\ell,v_k),(v_k,v_{k+1}),\ldots,(v_{k+q-1},$ $v_{k+q})$ are included
in $T$.

In Figure~\ref{canonical-triconnected}, the edges in $T$ are indicated
by the thick lines. By an argument similar to the proof of 
Lemma~\ref{lemma:3-tree}(\ref{lemma:3-tree_1}), $T$ is a spanning tree
of $G$.  As in \S\ref{sec_triangle}, we similarly define the {\em
rightmost} depth-first search in $T$.  Note that the order in which
the vertices of $T$ are visited by the rightmost depth-first search is
the rightmost canonical ordering $v_1,\ldots,v_n$ that defines $T$.

We are now ready to describe the encoding $S$ of $G$ by means of $T$.
We further divide Case 1 into three subcases.

{\em Case 1a}: No vertex is covered at step $k'$.

{\em Case 1b}: At least one vertex is covered at step $k'$ and the leftmost
covered vertex $c_{\ell+1}$ is adjacent to $v_k$.

{\em Case 1c}: At least one vertex is covered at step $k'$ and the leftmost
covered vertex $c_{\ell+1}$ is not adjacent to $v_k$.

Let $\beta$ be the number of steps for reconstructing $G$.
Let $\beta_{1a},\beta_{1b},\beta_{1c}$ and $\beta_2$ be the numbers of
steps of Cases 1a, 1b, 1c, and 2, respectively.  We first consider the case
$\beta_{1b} \geq \beta_{1c}$ to encode $G$ with {\em Scheme I}; afterwards, we
modify Scheme I into {\em Scheme II} for the case $\beta_{1b} < \beta_{1c}$.

In Scheme I, the encoding $S$ of $G$ is the concatenation of three strings
$S_1$, $S_2$ and $S_3$. $S_1$ is the folklore parentheses encoding of $T$,
which is rooted and ordered in the same way as in \S\ref{sec_triangle}.
Since $T$ has $n$ vertices, $S_1$ has $2(n-1)$ bits.

To construct $S_2$, first let $Q=s_1 * s_2 * \cdots * s_\beta *$ where
each $s_{k'}$ is a binary string that corresponds to the step $k'$ 
of reconstructing $G$ based on the ordering $v_1,\ldots,v_n$.
$s_{k'}$ is determined as follows. 
The following two cases both assume that $d$ vertices are covered at
step $k'$.

Case 1. Note that $d=r-\ell-1$.  The string $s_{k'}$ has $d$ symbols
corresponding to $c_j$ with $j=\ell+1,\ldots,r-1$, respectively. If the edge
$(c_j,v_k)$ is present in $G$, the symbol in $s_{k'}$ corresponding to $c_j$
is 1; otherwise, the symbol is 0. Note that in Case 1a, since no vertex is
covered, $s_{k'}$ is empty.

Case 2.  The string $s_{k'}$ consists of $q$ copies of $0$ followed by $d$
copies of $1$.

For example, the string $Q$ for Figure \ref{canonical-triconnected} is:

\begin{center} 
$\begin{array}{cccccccccccccc} \underbrace{00} & * & 0 & * & * & 0 & *
& 0 & * & \underbrace{1 0 0 0} & * & * & \underbrace{1 0 0 0 1} & *
\nonumber \\
\uparrow &  &\uparrow && & \uparrow && \uparrow & & \uparrow &&& \uparrow & \nonumber \\
s_1 &  & s_2 &  & & s_4 && s_5 & & s_6 && & s_8 & \nonumber \\
\end{array}$
\end{center}

$S_2$ is a binary representation of $Q$ defined as follows.  A step of Case
1 adds one vertex to $G$ and correspondingly includes one $*$ in $Q$;
similarly, a step of Case 2 adds $q+1$ vertices to $G$ and includes one $*$
and $q$ copies of $0$ in $Q$. Since exactly $n-2$ vertices are added, the
total number of these symbols is $n-2$.  Each symbol in $Q$ not yet counted
corresponds to a vertex covered at the $\beta$ steps. Since each $v_k$ with $3
\leq k \leq n-1$ is covered at most once and $v_1,v_2,v_n$ are never
covered, the total number of these latter symbols is at most $n-3$. Thus
$Q$ has at most $2n-5$ symbols.  For the sake of unambiguous decoding, we
pad $Q$ with copies of 1 at its end to have exactly $2n-5$ symbols.  Since
$Q$ uses 3 distinct symbols, we treat it as an integer of base 3 and
convert it to a binary integer.  Again, for the sake of unambiguous
decoding, we use exactly $\lceil{(2n-5)\log{3}}\rceil$ bits for this binary
integer by padding copies of $0$ at its beginning. The resulting binary
string is the desired $S_2$.

For the sake of decoding, we also need to know whether any given $s_{k'}$ is
of Case 1 or 2. Thus, let $S_3=t_1 \cdots t_\beta$ where $t_{k'}=1$ if step $k'$
is of Case 1 and $t_{k'}=0$ otherwise. To save space, note that some bits
$t_{k'}$ can be deleted as follows without incurring ambiguity. If step $k'$
is of Case 1a, $t_{k'}$ is deleted because $s_{k'}$ is empty and only a string of
Case 1a can be empty. If step $k'$ is of Case 1b, $t_{k'}$ is deleted because
$s_{k'}$ starts with 1, while the strings of Case 2 start with 0. If step $k'$
is of Case 1c or 2, $t_{k'}$ remains in $S_3$. For example, the string $S_3$
for Figure~\ref{canonical-triconnected} consists of $t_1=0$, $t_2=0$,
$t_4=1$, $t_5=0$.  Thus, $S_3$ has $\beta_{1c}+\beta_2$ bits, which can be
bounded as follows.  A step of Case 1 adds one vertex into $G$ and a step
of Case 2 adds at least two vertices. Since $n-2$ vertices are added over
the $\beta$ steps, $\beta_{1a}+\beta_{1b}+\beta_{1c}+2 \beta_2 \leq n-2$.  Since Scheme I
assumes $\beta_{1b} \geq \beta_{1c}$, $|S_3| = \beta_{1c}+\beta_2 \leq
\frac{1}{2}{\cdot}(\beta_{1b}+\beta_{1c}) + \beta_2 \leq
\frac{1}{2}{\cdot}(\beta_{1a}+\beta_{1b}+\beta_{1c}+2\beta_2) \leq 0.5n -1$.

Since $S= S_1 // S_2// S_3$, $|S| \leq 2(n-1)+\lceil(2n-5)\log
3\rceil+0.5n-1 \leq (2.5+2\log 3)n - 9$ bits. This completes the
description of the encoding procedure of Scheme I.

Next we describe how to decode $S$ to reconstruct $G$.  This decoding
assumes that both $S$ and $n$ are given. Thus, we can uniquely determine
$S_1$, $S_2$ and $S_3$. Then we convert $S_2$ to $Q$.  {From} $Q$ we can
recover all $s_{k'}$ with $1\leq k' \leq \beta$. {From} $S_3$ and all $s_{k'}$, we can
recover all $t_{k'}$ with $1\leq k' \leq \beta$. {From} $S_1$, we reconstruct
$T$. From $T$, we find the ordering $v_1,\ldots,v_n$.  Afterwards, we draw
the edge $(v_1,v_2)$ and perform a loop of steps as follows. Each step is
indexed by $k'$ and corresponds to step $k'$ of reconstructing $G$ using the
rightmost canonical ordering.

If $t_{k'}=1$, step $k'$ is of Case 1. Thus, a vertex $v_k$ is added at this
step where $v_k$ is the smallest ordered vertex not added into the current
graph yet. {From} $T$, we can determine the leftmost neighbor $c_\ell$ of $v_k$ in
the contour $H_{k-1}$ because $c_\ell$ is the parent of $v_k$ in $T$.  {From}
$s_{k'}$, we know the number of vertices covered by $v_k$ and hence the
rightmost neighbor $c_r$ of $v_k$ in the contour $H_{k-1}$. {From} $s_{k'}$,
we also know which of the covered vertices are connected to $v_k$. These
corresponding edges are added to $G$.

If $t_{k'}=0$, step $k'$ is of Case 2. Thus, a chain $v_k,\ldots,v_{k+q}$ is
added at this step where $v_k$ is the smallest ordered vertex not added into
the current graph yet. The integer $q$ can be determined from the string $s_{k'}$
by counting its leading copies of $0$. {From} $s_{k'}$, we also know the
number of vertices covered at step $k'$, which is the count of $1$ in $s_{k'}$.
Thus, we know the neighbor $c_r$ of $v_{k+q}$ in the contour $H_{k-1}$. The
chain is added accordingly.

This completes the decoding procedure of Scheme I. It is straightforward to
implement the whole Scheme I in $O(n)$ time.  If $\beta_{1b} < \beta_{1c}$, we use
Scheme II to encode $G$, which is identical to Scheme I with the following
differences.  If step $k'$ is of Case 2, $s_{k'}$ consists of $q$ copies of 1
followed by $d$ copies of 0. Also, all bits $t_{k'}$ for steps of Cases 1a and
1c are omitted from $S_3$ without incurring ambiguity  since their corresponding
strings $s_{k'}$ either are empty or start with 0 while the strings of Cases
1b and 2 start with 1.  We use one extra bit to encode whether we use
Scheme I or II. Thus we have the following lemma.

\begin{lemma}\label{encode-triconnected}
  Any triconnected plane graph with $n$ vertices can be encoded using at
  most $(2.5+2\log 3)n - 8$ bits. Both encoding and decoding take $O(n)$
  time. The decoding procedure assumes that both $S$ and $n$ are given.
\end{lemma}

We can improve Lemma~\ref{encode-triconnected} as follows.  Let $G^*$ be
the dual of $G$. $G^*$ has $f$ vertices, $m$ edges and $n$ faces.  Since
$G$ is triconnected, $G^*$ is also triconnected. Furthermore, if $n > 3$,
then $f > 3$ and $G^*$ has no self-loop or multiple edge. Thus, we can use
the coding scheme of Lemma~\ref{encode-triconnected} to encode $G^*$ with
at most $(2.5+2\log 3)f - 8$ bits.  Since $G$ can be uniquely determined
from $G^*$, to encode $G$, it suffices to encode $G^*$. To make $S$
shorter, for the case $n > 3$, if $n \leq f$, we encode $G$ using at most
$(2.5+2\log 3)n - 8$ bits; otherwise, we encode $G^*$ using at most
$(2.5+2\log 3)f - 8$ bits. This new encoding has at most
$(2.5+2\log{3})\min\{n,f\}-8$ bits. Since $\min\{n,f\} \leq
\frac{n+f}{2}$, the bit count is at most $(1.25+\log 3)m-2$ by Euler's
formula $n+f=m+2$.  For the sake of decoding, we use one extra bit to
denote whether we encode $G$ or its dual. Note that if $n =3$, we can
simply encode $G$ using zero bit without ambiguity.  Thus we have proved
the following theorem.

\begin{theorem}\label{encode-triconnected-2}
  Any triconnected plane graph with $n$ vertices, $m$ edges and $f$ faces
  can be encoded using at most $(2.5+2\log{3})\min\{n,f\}-7 \leq
  (1.25+\log{3})m-1$ bits. Both encoding and decoding take $O(n)$ time. The
  decoding procedure assumes that $S$ is given together with $n$ or $f$ as
  appropriate.
\end{theorem}

{\it Remark.} There are several ways to improve this coding scheme so that
the decoding does not require $n$ as input. One is to use well-known data
compression techniques to encode $n$ and append it to the beginning of $S$
using $\log n + O(\log\log n)$ bits \cite{BCW90, Elias75}.  Another is to
pad $S$ with copies of $1$ at its end so that it has exactly
$\lceil(2.5+2\log{3})\min\{n,f\}\rceil-7$ bits. Then, since $2.5+2\log{3} >
1$, given $S$ alone, we can uniquely determine $n$ or $f$ and proceed with
the original decoding procedure.  With the strings $s_{k'}$, we can
unambiguously identify the padded bits.

\section{Open Problems}\label{sec_open}
This paper leaves several problems open. Since plane triangulations
are useful in many application areas, it would be particularly helpful
to encode them in $O(n)$ time using close to $1.08m$ bits. Similarly,
it would be significant to obtain a linear-time coding scheme for
triconnected plane graphs using close to $2m$ bits. Note that Tutte
\cite{Tutte63b} proved an information-theoretic tight bound of
$2m+o(m)$ bits for triconnected plane graphs that may contain multiple
edges and self-loops.  More generally, it would be of interest to
encode graphs in a given family in polynomial time using their
information-theoretic minimum number of bits.  Solving these problems
will most likely lead to the discovery of new structural properties of
graphs.

{\bf Acknowledgments.} The authors are grateful to anonymous referees
for helpful comments.


\begin{thebibliography}{10}

\bibitem{BCW90}
{\sc T.~C. Bell, J.~G. Cleary, and I.~H. Witten}, {\em Text Compression},
  Prentice-Hall, Englewood Cliffs, NJ, 1990.

\bibitem{Ber85}
{\sc C.~Berge}, {\em Graphs}, North-Holland, New York, NY, second revised~ed.,
  1985.

\bibitem{clr}
{\sc T.~H. Cormen, C.~L. Leiserson, and R.~L. Rivest}, {\em Introduction to
  Algorithms}, {MIT} Press, Cambridge, MA, 1990.

\bibitem{DeFPP90}
{\sc H.~de~Fraysseix, J.~Pach, and R.~Pollack}, {\em How to draw a planar graph
  on a grid}, Combinatorica, 10 (1990), pp.~41--51.

\bibitem{Elias75}
{\sc P.~Elias}, {\em Universal codeword sets and representations of the
  integers}, IEEE Transactions on Information Theory, IT-21 (1975),
  pp.~194--203.

\bibitem{GW83}
{\sc H.~Galperin and A.~Wigderson}, {\em Succinct representations of graphs},
  Information and Control, 56 (1983), pp.~183--198.

\bibitem{Har72}
{\sc F.~Harary}, {\em Graph Theory}, Addison-Wesley, Reading, MA, 1972.

\bibitem{IR82}
{\sc A.~Itai and M.~Rodeh}, {\em Representation of graphs}, Acta Informatica,
  17 (1982), pp.~215--219.

\bibitem{Jacobson89}
{\sc G.~Jacobson}, {\em Space-efficient static trees and graphs}, in
  Proceedings of the IEEE Thirtieth Annual Symposium on Foundations of Computer
  Science, 1989, pp.~549--554.

\bibitem{KNR92}
{\sc S.~Kannan, M.~Naor, and S.~Rudich}, {\em Implicit representation of
  graphs}, {SIAM} Journal on Discrete Mathematics, 5 (1992), pp.~596--603.

\bibitem{Kant92}
{\sc G.~Kant}, {\em Drawing planar graphs using the $lmc$-ordering}, in
  Proceedings of the~33rd Annual IEEE Symposium on Foundations of Computer
  Science, 1992, pp.~101--110.

\bibitem{KH97}
{\sc G.~Kant and X.~He}, {\em Regular edge labeling of 4-connected plane graphs
  and its applications in graph drawing problems}, Theoretical Computer
  Science, 172 (1997), pp.~175--193.

\bibitem{kaofhr94}
{\sc M.~Y. Kao, M.~F\"urer, X.~He, and B.~Raghavachari}, {\em Optimal parallel
  algorithms for straight-line grid embeddings of planar graphs}, {SIAM}
  Journal on Discrete Mathematics, 7 (1994), pp.~632--646.

\bibitem{kaot93.joa}
{\sc M.~Y. Kao, N.~Occhiogrosso, and S.~H. Teng}, {\em Simple and efficient
  compression schemes for dense and complement graphs}, Journal of
  Combinatorial Optimization,  (1999).
\newblock To appear.

\bibitem{KW:encodings}
{\sc K.~Keeler and J.~Westbrook}, {\em Short encodings of planar graphs and
  maps}, Discrete Applied Mathematics, 58 (1995), pp.~239--252.

\bibitem{Lovasz_book_86}
{\sc L.~Lov\'{a}sz}, {\em An Algorithmic Theory of Numbers, Graphs and
  Convexity}, Society for Industrial and Applied Mathematics, Philadelphia, PA,
  1986.

\bibitem{MR97}
{\sc J.~I. Munro and V.~Raman}, {\em Succinct representation of balanced
  parentheses, static trees and planar graphs}, in Proceedings of the 38th
  Annual IEEE Symposium on the Foundations of Computer Science, 1997,
  pp.~118--126.

\bibitem{naor90}
{\sc M.~Naor}, {\em Succinct representations of general unlabeled graphs},
  Discrete Applied Mathematics, 28 (1990), pp.~303--307.

\bibitem{PH86.encode}
{\sc C.~H. Papadimitriou and M.~Yannakakis}, {\em A note on succinct
  representations of graphs}, Information and Control, 71 (1986), pp.~181--185.

\bibitem{Schnyder90}
{\sc W.~Schnyder}, {\em Embedding planar graphs on the grid}, in Proceedings of
  the 1st Annual ACM-SIAM Symposium on Discrete Algorithms, 1990, pp.~138--148.

\bibitem{turan84}
{\sc G.~Tur\'{a}n}, {\em On the succinct representation of graphs}, Discrete
  Applied Mathematics, 8 (1984), pp.~289--294.

\bibitem{Tutte62}
{\sc W.~T. Tutte}, {\em A census of planar triangulations}, Canadian Journal of
  Mathematics, 14 (1962), pp.~21--38.

\bibitem{Tutte63b}
\leavevmode\vrule height 2pt depth -1.6pt width 23pt, {\em A census of planar
  maps}, Canadian Journal of Mathematics, 15 (1963), pp.~249--271.

\end{thebibliography}

\end{document}